\documentclass[conference]{IEEEtran}
\IEEEoverridecommandlockouts

\usepackage{geometry}   
\usepackage{mathrsfs}
\usepackage{amsmath}
\usepackage{amssymb}
\usepackage{stfloats}
\usepackage{graphicx}
\usepackage{setspace}
\usepackage{xcolor}
\usepackage{cite}
\usepackage{algorithm}      
\usepackage{algorithmic}    
\usepackage{multirow}       

\usepackage{tabularx,ragged2e,booktabs,caption}

\usepackage{fancyhdr}

\ifCLASSOPTIONcompsoc
\usepackage[tight,normalsize,sf,SF]{subfigure}
\else
\usepackage[tight,footnotesize]{subfigure}
\fi

\makeatletter
\long\def\@makecaption#1#2{\ifx\@captype\@IEEEtablestring%
\footnotesize\begin{center}{\normalfont\footnotesize #1}\\
{\normalfont\footnotesize\scshape #2}\end{center}%
\@IEEEtablecaptionsepspace
\else
\@IEEEfigurecaptionsepspace
\setbox\@tempboxa\hbox{\normalfont\footnotesize {#1.}~~ #2}%
\ifdim \wd\@tempboxa >\hsize%
\setbox\@tempboxa\hbox{\normalfont\footnotesize {#1.}~~ }%
\parbox[t]{\hsize}{\normalfont\footnotesize \noindent\unhbox\@tempboxa#2}%
\else
\hbox to\hsize{\normalfont\footnotesize\hfil\box\@tempboxa\hfil}\fi\fi}
\makeatother

\begin{document}
\title{Machine Learning-Based Antenna Selection in Untrusted Relay Networks
\IEEEauthorrefmark{1}\thanks{\IEEEauthorrefmark{1}
This work was supported in part by the National Natural Science Foundation of China (No. 61501376, 61871327 and 61801218), the Natural Science Foundation of Jiangsu Province (No. BK20180424), the Fundamental Research Funds for the Central Universities (No. 3102018JGC006), and the Aeronautical Science Foundation of China (2017ZC53029).
}
\thanks{\IEEEauthorrefmark{2}Corresponding author: yaorg@nwpu.edu.cn}
}

\author{\IEEEauthorblockN{Rugui Yao\IEEEauthorrefmark{2}\IEEEauthorrefmark{3},
Yuxin Zhang\IEEEauthorrefmark{3},
Nan Qi\IEEEauthorrefmark{4},
Theodoros A. Tsiftsis\IEEEauthorrefmark{5}
}
\IEEEauthorblockA{\IEEEauthorrefmark{3}School of Electronics and Information, Northwestern Polytechnical University, Xi'an 710072, Shaanxi, China}
\IEEEauthorblockA{\IEEEauthorrefmark{4}Department of Electronic Engineering, Nanjing University of Aeronautics and Astronautics, Nanjing 210016, China}
\IEEEauthorblockA{\IEEEauthorrefmark{5}School of Electrical and Information Engineering, Jinan University (Zhuhai Campus), Zhuhai 519070, China}
}

\maketitle

\begin{abstract}
This paper studies the transmit antenna selection based on \emph{machine learning} (ML) schemes in untrusted relay networks.
First, we state the conventional antenna selection scheme.
Then, we implement three ML schemes, namely, the support vector machine-based scheme, the naive-Bayes-based scheme, and the k-nearest neighbors-based scheme, which are applied to select the best antenna with the highest secrecy rate.
The simulation results are presented in terms of system secrecy rate and secrecy outage probability.
From the simulation, we can conclude that the proposed ML-based antenna selection schemes can achieve the same performance without amplification at the relay, or small performance degradation with transmitted power constraint at the relay, comparing with conventional schemes. However, when the training is completed, the proposed schemes can perform the antenna selection with a small computational complexity.
\end{abstract}


\IEEEpeerreviewmaketitle

\section{introduction}

Nowadays, wireless communication network has become an indispensable part of civilian life and military applications. In order to meet the demand for high quality communication services, there are many emerging wireless technologies such as massive MIMO, millimeter wave communications and machine type communication, etc. \cite{Talwar2014Enabling, Yang2015Safeguarding, Wu2018A}. In order to ensure private and important information transmission, physical layer security is the state-of-the-art technique for future wireless networks. \emph{Transmit antenna selection} (TAS) is an efficient scheme to guarantee the high spectral efficiency and overcome the hardware energy consumption in communication environments\cite{Ahmed2017A, Joung2016Two}.

On the other hand, in recent years, \emph{machine learning}(ML) schemes have attracted a lot of attention for applications in wireless communications, such as \emph{support vector machine} (SVM) \cite{Ertekin2011Nonconvex}, \emph{k-nearest neighbors} (k-NN) \cite{Chang2011Semi} and \emph{naive-Bayes} (NB) \cite{Netti2016A}, which demonstrate the superior performance on multiclass classification. ML schemes can be combined with TAS to classify the \emph{channel state information} (CSI) and obtain the best selected antenna which can achieve the highest system secrecy rate and reduce computation of complexity.

Cooperative jamming\cite{Tang2015Social}, cooperative relay \cite{Li2011On} and other jamming schemes have been widely employed in wiretap communication networks, since it helps protecting the networks from eavesdropper attacks and guaranteeing the secure communication between the source and the destination, thereby improving the physical layer security.

In this paper, we apply the ML-based TAS in the untrusted relay network, aiming to enhance the physical layer security and reducing the computational complexity. We first construct the system model and signal model with consideration of TAS. Then, the ML-based TAS for the untrusted relay network is developed, where k-NN, NB and SVM algorithms are considered. Finally, the simulated secrecy rate, \emph{secrecy outage probability} (SOP) and misclassification rate for the conventional transverse algorithm and ML-based algorithm are presented and compared.

\section{System and Signal Model}\label{signal Model}
\begin{figure}[ht]
    \centering
    \includegraphics[width=0.45\textwidth] {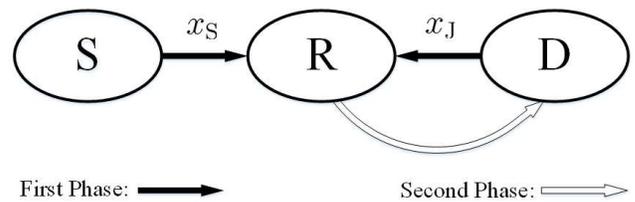}
    \caption{cooperative jamming in untrusted relay network}\label{system_model}
\end{figure}

As shown in Fig.\ref{system_model}, we consider cooperative jamming in untrusted relay networks, which consists of a source (S), a destination (D) and an untrusted \emph{amplifying-and-forward} (AF) relay (R). All the nodes, S, R, and D, are equipped with $N_\mathrm{S}$, $N_\mathrm{R}$, $N_\mathrm{D}$ antennas, respectively. For simplicity, we assume $N_\mathrm{R} = N_\mathrm{D} =1$ for our initial work. In the untrusted relay network, we assume the relay is untrustworthy, and works in \emph{time-division multiple-access }(TDMA) mode and a half-duplex two-hop relaying. Only the source S employed the TAS schemes. In addition, all channels are subject to \emph{independent and identically distributed} (i.i.d) Rayleigh fading.

In this model, due to shadowing or too long distance, there is no direct link between S and D, thus the communication between S and D is implemented via R.
We denote $\mathbf{h} = \left[h_1, \cdots, h_{N_\mathrm{S}}\right] \in \mathcal{C}^{1 \times N_\mathrm{S}}$ as the channel vector from S to R. Further define $g_\mathrm{R-D} \in \mathcal{C}^{1 \times 1}$ and $g_\mathrm{D-R} \in \mathcal{C}^{1 \times 1}$  as the channel gains from R to D and from D to R. Here, since the channel reciprocity is considered in this paper, we have $g_\mathrm{R-D} = g_\mathrm{R-D}^*$. And let $g_\mathrm{R-D} = g$ for simple representation.
As regarding the high cost of RF chain, in this system, only $N_\mathrm{T}$ antennas among $N_\mathrm{S}$ of S are activated to perform transmission.
Assume the available $N_\mathrm{S}$ antennas are labeled as $1, 2, \cdots, N_\mathrm{S}$ and the selected $N_\mathrm{T}$ antennas are with the indices $s_1,s_2, \cdots, s_{N_\mathrm{T}}$ where $s_j \in [1, N_\mathrm{S}]$ for $j=1, \cdots, N_\mathrm{T}$. Therefore, the practical propagation channel from S to R can be denoted as $\widetilde{\mathbf{h}}=\left[h_{s_1}, \cdots, h_{s_{N_\mathrm{T}}}\right] \in \mathcal{C}^{1 \times N_\mathrm{T}}$. In order to maximize the received SNR for the relay R and thus the destination D, matched filter precoding is applied. In this case, the precoding vector for S's transmission is $\mathbf{p}_\mathrm{MF} = \frac{\widetilde{\mathbf{h}}^{\mathrm{H}}}{\|\widetilde{\mathbf{h}}\|_2}$, where $\|\cdot\|_2$ represents the $2$ norm of a vector.


Owing to the untrusted relay, we adopted the \emph{destination-aided jamming} (DAJ) technique\cite{yao_access,yao_tifs} and implement the transmission in two time-slots.

In the first time slot, S transmits its precoded signal, $\mathrm{P}_\mathrm{MF} x_\mathrm{S}$, and meanwhile D transmits cooperative jamming signal $x_\mathrm{J}$ to the untrusted relay R. The received signal at R, $y_\mathrm{R}$, is given by \cite{Yao2018Optimal,yao_tifs}

\begin{align}\label{y_R}
y_\mathrm{R} &= \sqrt{\frac{P_\mathrm{S}}{N_\mathrm{T}}} \widetilde{\mathbf{h}} \mathbf{p}_\mathrm{MF} x_\mathrm{S} + \sqrt{P_\mathrm{D}} g_\mathrm{D-R}x_\mathrm{J} + n_\mathrm{R}\nonumber\\
&=\sqrt{\frac{P_\mathrm{S}}{N_\mathrm{T}}} \|\widetilde{\mathbf{h}}\|_2 x_\mathrm{S} + \sqrt{P_\mathrm{D}} g x_\mathrm{J} + n_\mathrm{R},
\end{align}
\noindent where $x_\mathrm{S}$ and $x_\mathrm{J}$ denote the confidential signal and cooperative jamming signal with unit power, i.e., $|x_\mathrm{S}|=|x_\mathrm{J}|=1$;
$P_\mathrm{S}$ and $P_\mathrm{D}$ are the transmitted powers for confidential and cooperative jamming signals, respectively;
$n_\mathrm{R}$ denotes the complex \emph{additive white Gaussian noise} (AWGN) received at R, following $\mathcal{CN}(0, N_0)$-distribution. In this paper, we assume all the AWGNs received at R in the first time slot and at D in the second time slot are both with unit \emph{power spectral density} (PSD), that is, $N_0 = 1$. As a result, the \emph{signal-to-noise ratio} (SNR) at different nodes can be adjusted by the transmitted power. With the cooperative jamming signal as the second item in (\ref{y_R}), the interpretation of $x_\mathrm{S}$ by R is degraded.

From (\ref{y_R}), the instantaneous received \emph{signal-to-interference-plus-noise ratio} (SINR) at R can be presented as
\begin{align}\label{gamma_R}
\gamma_\mathrm{R} = \frac{\frac{P_\mathrm{S}}{N_\mathrm{T}} \|\widetilde{\mathbf{h}}\|_2^2
}
{\left(P_\mathrm{D} |g|^2 + 1\right)}.
\end{align}

In the second time slot, the relay R re-transmits the received signals to D after amplifying it with amplification factor $\beta$. Let $P_\mathrm{R}$ be the transmitted power by R. Therefore, with $y_\mathrm{R}$ in (\ref{y_R}), the amplification factor $\beta$ must satisfy the following equation
\begin{align}\label{beta2}
\beta^2 = \frac{P_\mathrm{R}}{\frac{P_\mathrm{S}}{N_\mathrm{T}}
\|\widetilde{\mathbf{h}}\|_2^2
+ P_\mathrm{D} |g|^2 + 1}.
\end{align}

Then the received signal at D from the untrusted relay is given by

\begin{align}\label{y_D}
y_\mathrm{D} &= \beta g_\mathrm{R-D} \frac{\sqrt{P_\mathrm{S}}}{N_\mathrm{T}}
\|\widetilde{\mathbf{h}}\|_2^2
x_\mathrm{S} + \beta g_\mathrm{R-D} \sqrt{P_\mathrm{D}} g_\mathrm{D-R}x_\mathrm{J} \nonumber\\
&~~~+ \beta g_\mathrm{R-D} n_\mathrm{R} + n_\mathrm{D},
\end{align}
\noindent where $n_\mathrm{D}$ is the complex AWGN received at D, which is also assumed to be $\mathcal{CN}(0, 1)$-distributed.

Since the second item in (\ref{y_D}) is transmitted by D itself, D can perform self-interference cancellation with perfect \emph{channel state information} (CSI) available. Consequently, the received signal at D in (\ref{y_D}) can be rewritten as
\begin{align}\label{y_D2}
y_\mathrm{D} = \beta g^* \frac{\sqrt{P_\mathrm{S}}}{N_\mathrm{T}} \|\widetilde{\mathbf{h}}\|_2^2 x_\mathrm{S} + \beta g^* n_\mathrm{R} + n_\mathrm{D}.
\end{align}

And from (\ref{y_D2}), the instantaneous SINR at D can be presented as
\begin{align}\label{gamma_D}
\gamma _\mathrm{D}=\frac{\frac{P_\mathrm{S}}{N_\mathrm{T}} \beta^2 \| \widetilde{\mathbf{h}}\|_2^2 | g |^2}{\beta^2 | g |^2 + 1}.
\end{align}

In the physical-layer security based untrusted relay system, the achievable secrecy rate can be defined as \cite{Yao2016Optimised, yao_tifs, yao_access}
\begin{align}\label{R_s}
R_s=\left[\log_2(1+\gamma_\mathrm{D})-\log_2(1+\gamma_\mathrm{R})\right]^+,
\end{align}
\noindent where $[\cdot]^+=\max(\cdot, 0)$. Note that, for simplification, we neglect this operator for the following derivation but consider it for simulation.

With (\ref{gamma_R}) and (\ref{gamma_D}), we can formulate the secrecy rate in (\ref{R_s}) after some calculation as
\begin{align}\label{R_s_all}
R_s&=\log_2 \left(
\frac{\frac{P_\mathrm{S}}{N_{\mathrm{T}}} \|\widetilde{\mathbf{h}}\|_2^2}{P_\mathrm{D}|g|^2+1}
\right.\nonumber\\
&\times \left.
\frac{\frac{P_\mathrm{S}}{N_{\mathrm{T}}} P_\mathrm{R} |g|^2 \|\widetilde{\mathbf{h}}\|_2^2
+P_\mathrm{R} |g|^2+\frac{P_\mathrm{S}}{N_\mathrm{T}}
\|\widetilde{\mathbf{h}}\|_2^2
+ P_\mathrm{D} |g|^2 + 1
}
{P_\mathrm{R}|g|^2+\frac{P_\mathrm{S}}{N_\mathrm{T}}
\|\widetilde{\mathbf{h}}\|_2^2
+ P_\mathrm{D} |g|^2 + 1}
\right).
\end{align}

From (\ref{R_s_all}) and considering $\|\widetilde{\mathbf{h}}\|^2_2=\sum_{j=1}^{N_\mathrm{T}} {|h_{s_j}|^2}$, $R_s$ has a very complicated coupling relationship with $\widetilde{\mathbf{h}}$ and $g$, which makes the antenna selection difficult to solve.

In the following part, we further consider three special and simplified scenarios.

\subsection{Special Case 1: No Amplification at R}
In this case, $\beta^2 = 1$; as a result, the achievable secrecy rate in (\ref{R_s_all}), $R_s$, can be simplified as
\begin{align}\label{R_s_case1}
R_s=\log_2 \left(
\frac{\frac{P_\mathrm{S}}{N_{\mathrm{T}}} \|\widetilde{\mathbf{h}}\|_2^2}{P_\mathrm{D}|g|^2+1}
\times
\frac{\frac{P_\mathrm{S}}{N_{\mathrm{T}}} |g|^2 \|\widetilde{\mathbf{h}}\|_2^2
+ |g|^2 + 1
}
{|g|^2+1}
\right).
\end{align}

In (\ref{R_s_case1}), only the nominator has relationship with $\widetilde{\mathbf{h}}$, which contributes to the correctness and simplification for the antenna selection.

\subsection{Special Case 2: Only One Selected Antenna}
In this case, we can select only one antenna, namely $h_s$, to implement the transmission. Therefore, $\|\widetilde{\mathbf{h}}\|^2_2$ can be further denoted as $|h_s|^2_2$. And the achievable secrecy rate in (\ref{R_s_all}), $R_s$, can be reformulated as
\begin{align}\label{R_s_case2}
R_s&=\log_2 \left(
\frac{P_\mathrm{S} |h_s|^2_2}{P_\mathrm{D}|g|^2+1}
\right.\nonumber\\
&\times \left.
\frac{P_\mathrm{S} P_\mathrm{R} |g|^2 |h_s|^2_2
+P_\mathrm{R} |g|^2+P_\mathrm{S}
|h_s|^2_2
+ P_\mathrm{D} |g|^2 + 1
}
{P_\mathrm{R}|g|^2+P_\mathrm{S}
|h_s|^2_2
+ P_\mathrm{D} |g|^2 + 1}
\right).
\end{align}

\subsection{Special Case 3: No Amplification at R and Only One Selected Antenna}
By combining the considerations in Special Cases 1 and 2, the $R_s$ in (\ref{R_s_all}) can be rewritten as
\begin{align}\label{R_s_case3}
R_s=\log_2 \left(
\frac{P_\mathrm{S} |h_s|^2_2}{P_\mathrm{D}|g|^2+1}
\times
\frac{P_\mathrm{S} |g|^2 |h_s|^2_2
+ |g|^2 + 1
}
{|g|^2+1}
\right).
\end{align}

\section{Conventional Antenna Selection Scheme}\label{con sch}
In conventional antenna selection scheme, the source S is aware of all CSIs, such as $\mathbf{h}$ and $g$. Then, S traverses all possible combinations for selected antennas, and computes the corresponding secrecy rate. The maximum secrecy rate and the corresponding antenna selection scheme are the solution for the consideration of this paper. The optimization problem can be formulated as
\begin{align}\label{opt_prob}
n^* = \arg \max_{n \in \mathcal{L}} R_s,
\end{align}
\noindent where $\mathcal{L}$ denote the index set for all possible combinations for selected antennas, with size $
\mathbb{C}_{N_\mathrm{S}}^{N_\mathrm{T}} =\frac{N_\mathrm{S}!}{N_\mathrm{T}!(N_\mathrm{S}-N_\mathrm{T})!}$.

Besides secrecy rate defined in (\ref{R_s}), SOP is another indicator to evaluate the system performance, which is defined as
\begin{align}\label{SOP}
P_\mathrm{out}(R_t) = \mathbb{P}(R_s < R_t),
\end{align}

\noindent where $\mathbb{P}(\cdot)$ is the probability, $R_t$ is the target SOP. In the simulation part, we will present the performance comparison for different antenna selection scheme with the two indicator, secrecy rate and SOP.

\section{Machine Learning-Based Transmit Antenna Selection Schemes}\label{ML-TAS}
In this section, we apply ML schemes, namely, SVM, NB and k-NN, into TAS scheme in an untrusted relay network. After feature vectors is generated, we construct the ML classification models. Based on the indicator in (\ref{R_s_all}), we can get class label of training CSI samples, and the class label can be as a index to select current best channel or channel combination. By inputting the training CSI samples and corresponding class labels to these ML models, we can predict the class labels of the new CSI samples.

\subsection{Construct training data sets}\label{construct data sets}
Before the ML models construction, we perform three-step procedure to construct training data sets \cite{Joung2016Machine}.

(1) \textbf{Preprocessing with Training Data Sets}: Preprocessing data sets are mainly to generate the input variables of ML system. Now, we obtain $M$ CSI samples as training data sets which is $\left[(\mathbf{h}_1^m,g_1^m),(\mathbf{h}_2^m,g_2^m), \cdots,(\mathbf{h}_M^m,g_M^m)\right]$. Then, we should do the preprocessing with the training data sets. The preprocessing process can be divided to 3 steps:

Step1: From (\ref{R_s_all})-(\ref{R_s_case3}), we can discover that the absolute values of channel gains of both $h_i$ and $g$ determine the antenna selection. Therefore, in this paper, to construct the feature vector, we generate a $1\times N$ real vector, $\mathbf{d}^m$, for $m\in \{1,\cdots, M\}$ and $N = N_\mathrm{S} + 1$, which can be denoted as
\begin{align}\label{vectors}
\mathbf{d}^m = \left[|h^m_1|, \cdots, |h^m_{N_\mathrm{S}}|, |g^m|\right],
\end{align}
\noindent where $h^m_i$ denote the $i$-th element of the $m$-th channel sample, $\mathbf{h}^m$, for $i=1,\cdots, N_\mathrm{S}$ and $m = 1, \cdots, M$.

Step2: Repeat Step1 for $M$ times, and we can get $M$ training feature vectors, i.e., $\mathbf{d}^1,\mathbf{d}^2, \cdots,\mathbf{d}^M$.

Step3: Generate $1 \times N$ normalized feature vector $\mathbf{t}^{m}$ by normalizing $\mathbf{d}^{m}$. The $i$-th element of $\mathbf{t}^{m}$, $t_i^m$, can be generated as
\begin{align}\label{normalize}
t_i^m = \frac{d_i^m -  \mathbb{E}[\mathbf{d}^m]}{\max(\mathbf{d}^m) -  \min(\mathbf{d}^m)},
\end{align}
\noindent where $d_i^m$ is the $i$-th element of $\mathbf{d}^{m}$.

(2) \textbf{KPI Design}: In this paper, maximizing system secrecy rate is the target. Therefore, we configure secrecy rate originally defined in (\ref{R_s_all}) as the \emph{key performance indicator} (KPI).

(3) \textbf{Labeling}: Calculate KPI of each antenna, then we get the $M$ channel labels as an index of target antenna that can maximize the KPI.

\subsection{Building Learning Models}\label{Building Learning Models}
With the normalized feature vectors in (\ref{normalize}) and its corresponding labels, we can build a learning model, whose input is the CSI samples and output is the classified label corresponding to the index of the selected antenna in the set $\mathcal{L}$. In this paper, we adopt three ML models, namely, SVM, NB, and k-NN.

(1) \textbf{SVM-Based Model}

The SVM is to construct a hyper-plane or many hyper-planes in a high dimensional space, which can be used for classification. The SVM has 2 classifier model construction methods, namely, OVR (one-vs.-rest) and OVO (one-vs.-one). In this paper, we adopt OVR method to construct the SVM model. And the logistic regression problem can be solved by the presented formula


\begin{align}\label{z_l}
{{\mathbf{w}_l}}&=\mathop{{\rm{argmin}}}\limits_{{\mathbf{w}_l}}C\sum\limits_{m=1}^M {[{b_l}[m]{g_1}({\mathbf{w}_l^\mathrm{T}} f({\mathbf{t}^m}))]}\nonumber\\
&~~~+ \mathop {{\rm{argmin}}}\limits_{{\mathbf{w}_l}} C\sum\limits_{m = 1}^M {[(1 - {b_l}[m]) \times {g_0}(\mathbf{w}_l^\mathrm{T}  f ({{\mathbf{t}^m}}))]} \nonumber\\
&~~~+  \left\| {{\mathbf{w}_l}} \right\|_2^2/2
\end{align}

\noindent where ${{\mathbf{w}}_l} \in {\mathbb{R}^{M \times 1}}$ , ${l} \in \{{1,2,...,{N_s}}\}$ is a learning parameter vector; The \emph{Radial Basis Function} (RBF) kernel has two parameters, $C$ and $ f{( {\mathbf{t}^m})}$. Choice of $C$ and $ f{( {\mathbf{t}^m})}$ plays an important role in the SVM¡¯s performance. C is a penalty factor, which trades off misclassification of training examples against simplicity of the decision surface.
${f}({{\mathbf{t}}^m}) \in {\mathbb{R}^{1 \times N}}$ is another kernel parameter which denotes how far the influence of a single training example reaches,
and its $\mathrm{n}$-th element can be expressed as ${f_\mathrm{n}}({\mathbf{t}^m}) = \exp( - \left\| {{\mathbf{t}^m} - {\mathbf{t}^n}} \right\|/(2{\sigma ^2}))$ with $\sigma$ representing the variance of $ f{( {\mathbf{t}^m})}$; if the m-th training CSI's label is equal to $l$, ${b_l}[m] = 1$, otherwise, ${b_l}[m] = 0$;  the ${g_k}(n) = {\mathrm{max}(( - 1)^k}n + 1,0)$ is the cost function.

After the $\mathbf{w}_l$ is obtained, the SVM's model can conduct antenna selection. We generate a set of new nomalized feature vectors $\mathbf{t}^m$ as described in Section IV-A and input them to the SVM's model. The current CSI label can be obtained by using $\mathbf{t}$ to replace $\mathbf{t}^m$, thus the $l^*$ antenna with largest ${\mathbf{w}}_l^\mathrm{T} f(\mathbf{t}) $ among all classes can be selected.

(2) \textbf{NB-Based Model}

The construction of NB-based model is based on conditional probability. In this paper, the NB classifier calculate the conditional probability for each feature vector $\mathbf{t}$ assigning to all label classes.

The posterior probability for the normalized feature vector $\mathbf{t}$ belonging to the class $l$ can be expressed as \cite{He2018Transmit}

\begin{equation}\label{NB}
{\mathbb{P(}}l{\mathbf{|t) = }}\frac{{{\mathbb{P(}}l{\mathbb{)P(\mathbf{t}|}}l{\rm{)}}}}{{{\mathbb{P(\mathbf{t})}}}}{\rm{ = }}\frac{{{\mathbb{P(}}l{\rm{)}}}}{{{\mathbb{P(\mathbf{t})}}}}\prod\limits_{{{n = 1}}}^{{N}} {{\mathbb{P(}}{{{t}}_{{n}}}{\rm{|}}l{\rm{)}}}
\end{equation}

\noindent where $l \in \{1,2,\cdots,N_\mathrm{S}\}$, $n \in \{1,2,\cdots,N\}$, $\mathbb{P}(\mathbf{t}|l)$ is the class-conditional probability of occurrence of the feature vector $\mathbf{t}$  given the class $l$, $\mathbb{P}(l)$ is prior probability, $\mathbb{P}(\mathbf{t})$ is the occurrence probability of feature vector $\mathbf{t}$, $\mathrm{t}_n$ is the n-th element of $\mathbf{t}$.
According to (\ref{NB}), we can obtain the antenna label which achieving the maximal posterier probability when the feature vector is $\mathbf{t}$.

We note that the prior probability $\mathbb{P}(\mathbf{t})$ is irrevelant with class label and $\mathbb{P}(l)$ is a constant for all classes, so $\mathbb{P}(\mathbf{t})$ and $\mathbb{P}(l)$ has no effect on antenna selection. As such, we can remove no influential terms. Thus, the (\ref{NB}) can be simplified in order to obtain maximal posterier occurrence probability of feature vector $\mathbf{t}$. This formula is presented as

\begin{equation}\label{NB_simplify}
{l^*} = \mathop {{\rm{argmax}}}\limits_{l \in {\rm{\{ 1}},{\rm{2}},...,{{{N}}_{\mathrm{S}}}{\rm{\} }}} \prod\limits_{{{n = 1}}}^{{N}} {{\mathbb{P}}({{{t}}_{{n}}}{\rm{|}}l)}
\end{equation}

\noindent Where ${l^*}$ reprents the selected antenna. We note that the NB-based model can conduct antenna selection after ${\rm{P(}}{t}_{{n}}{|l)}$ is obtained. The NB model will output a antenna label after inputting a new CSI feature vector $\mathbf{t}$.

(3) \textbf{k-NN Based Model}

We note that we get the CSI samples and their corresponding labels before model construction. The k-NN classifier calculates the Euclidian distance between new channel feature vectors and the training channel feature vectors. The k-NN classifier assign the label to the new CSI feature vectors if there is a minimal Euclidian distance between the new and training channel. The Euclidean distance can be expresssed as

\begin{align}\label{Euclidean distance}
\mathbb{D}(\mathbf{t},\mathbf{t}^m)=\left\| \mathbf{t}^m - \mathbf{t} \right\|_2
\end{align}

\noindent where the $\mathbf{t}^{m}$ is the training normalized CSI feature vector, $\mathbf{t}$ is the new CSI feature vector.

According to (\ref{Euclidean distance}), we can obtain the new label of CSI feature vector which has corresponding minimal Euclidean distance.

\section{Simulation Results And Analysis}\label{simulation}
In this section, we present some simulation results to verify the efficiency of the ML-based schemes.
We set the training and testing sets for $M = 10000$ channel implementations. The source S is configured with $N_\mathrm{S} = 6$ antennas. In this simulation, $N_\mathrm{T} = 1$ or $2$ antennas will be selected out of the available $N_\mathrm{S} = 6$ antennas. For simplicity, the confidential and cooperative jamming signals are transmitted with the same power, that is, $P_\mathrm{S} =P_\mathrm{D}$. When considering amplification at R, we further have $P_\mathrm{S} =P_\mathrm{D} =P_\mathrm{R}$; otherwise, when $\beta^2=1$, no constraint is preformed on $P_\mathrm{R}$.

Note that, in the following simulation results, we use solid lines and dot-dash lines to present the single and two antenna selection cases, respectively.

\subsection{System Performance}

\begin{figure}[ht]
    \centering
    \includegraphics[width=0.45\textwidth] {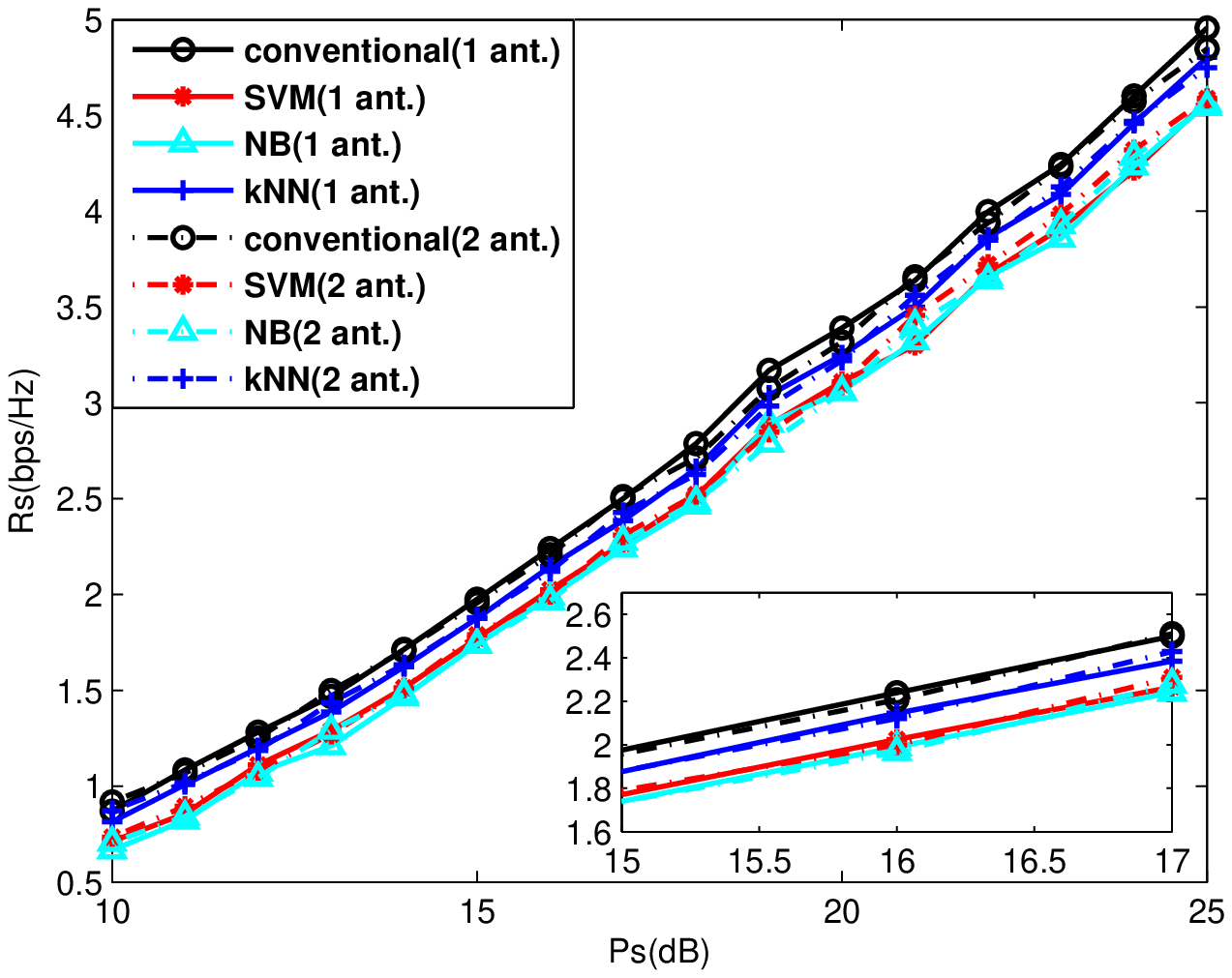}
    \caption{Secrecy rate of single and two antenna selection for $\beta^2  \ne 1$}\label{rate_beta}
\end{figure}

Fig.~\ref{rate_beta} shows the achievable secrecy rate of different single or two antenna selection schemes at different SNRs for $\beta^2 \ne 1$.
From Fig.~\ref{rate_beta}, we can discover that the secrecy rate increases for all schemes as the SNR increases.
The secrecy rate achieved by ML-based schemes has the same trend as that achieved by conventional scheme but with a small degradation. From (\ref{R_s_all}) and (\ref{R_s_case2}), the achievable secrecy rate, $R_s$, has a complicated coupling relationship with the selected channel gain $\widetilde{\mathbf{h}}$, which is difficult for the considered ML algorithm to thoroughly decouple. As a result, some misclassifications are emerged. This fact contributes the degradation. For decoupling the complicated relationship, we will employ Deep Learning model to shrink the gap as our future work.
Then, consider the number of selected antennas, the secrecy rate for single antenna selection is higher than that for two antenna selection. It is because that for single antenna selection, all power possessed by S is used to transmit the confidential signal through the only strongest channel. Whereas, for two antenna selection, half of the power is shared to the second strongest channel, which has smaller gain than the strongest one.

\begin{figure}[ht]
    \centering
    \includegraphics[width=0.45\textwidth] {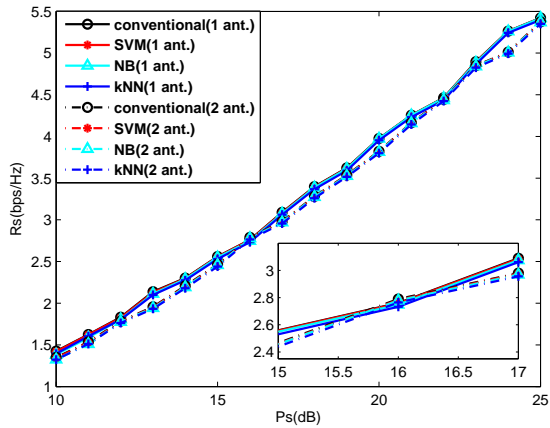}
    \caption{Secrecy rate of single and two antenna selection for $\beta^2=1$}\label{rate_beta_1}
\end{figure}

In Fig.~\ref{rate_beta_1}, we further plot the achievable secrecy rate of different single or two antenna selection schemes at different SNRs for $\beta^2 = 1$.
Comparing the results in Figs.~\ref{rate_beta} and ~\ref{rate_beta_1}, the achievable secrecy rate for $\beta^2 = 1$ shares the same trend with that for $\beta^2 \ne 1$. However, the gap between conventional and ML-based TAS schemes is diminished.
Considering the secrecy rate in (\ref{R_s_all}), (\ref{R_s_case1}) and (\ref{R_s_case3}), the selected channel gains just have impact on the nominator for $\beta^2 = 1$. This significantly reduce the complexity and couple, which make the ML-based TAS scheme predict the best selection precisely. Consequently, the ML-based TAS schemes achieve the same secrecy rate as conventional scheme.

\begin{figure}[ht]
    \centering
    \includegraphics[width=0.45\textwidth] {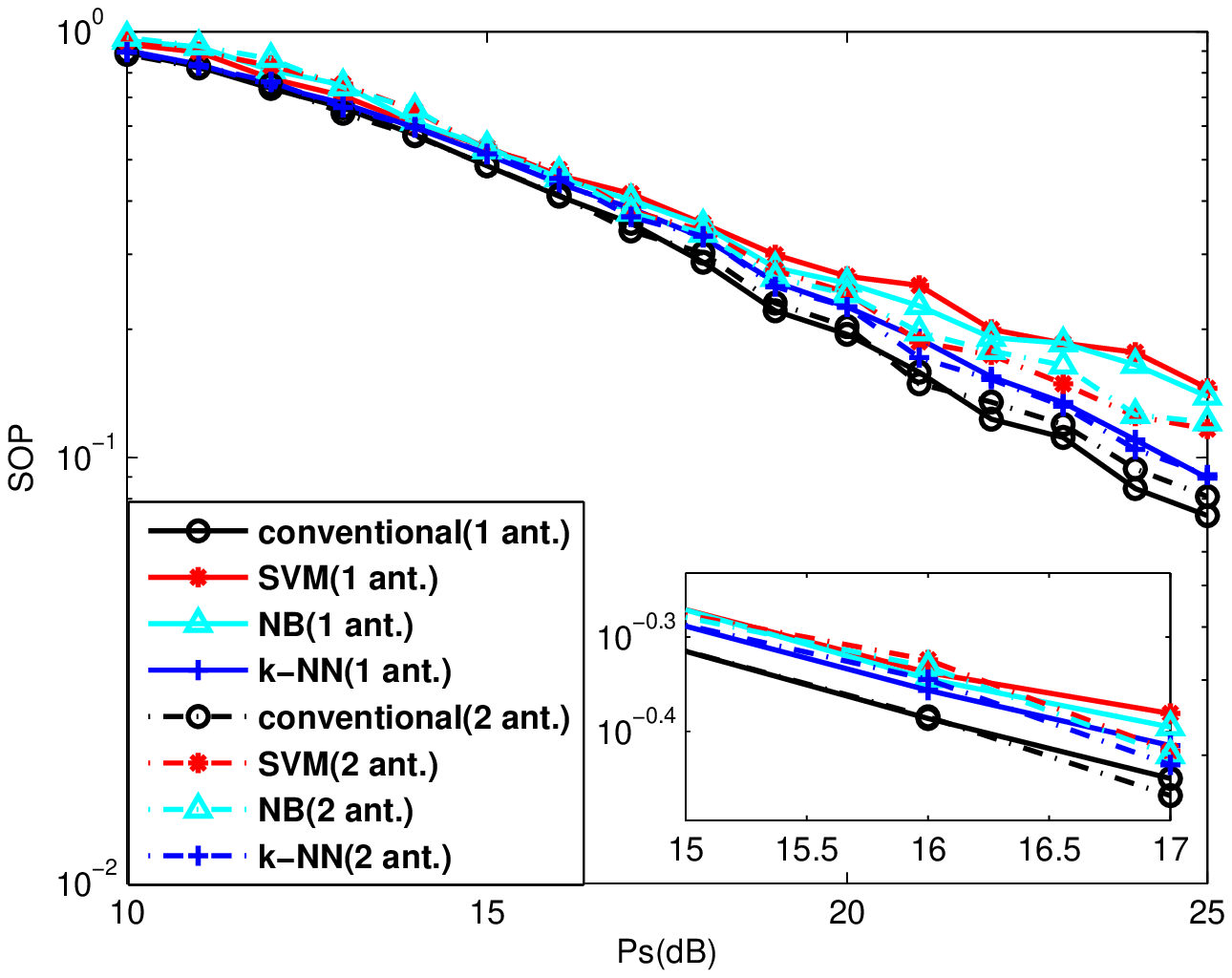}
    \caption{SOP of single and two antenna selection for $\beta^2 \ne 1$ and $R_t$ = 2 bps/Hz. }  \label{sop_beta}
\end{figure}

\begin{figure}[ht]
    \centering
    \includegraphics[width=0.45\textwidth] {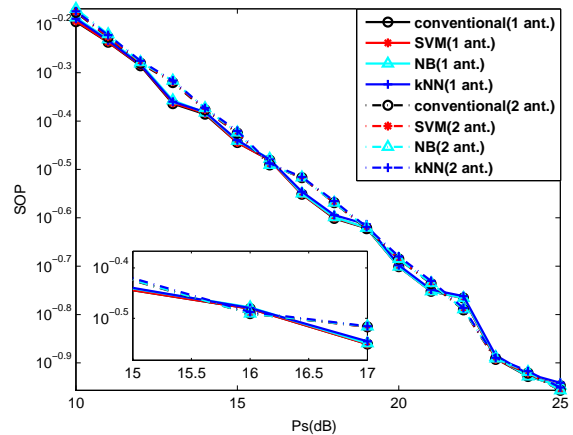}
    \caption{The SOP of single and two antennas selection for $\beta^2 = 1$ and $R_t$ = 2 bps/Hz}\label{sop_beta_1}
\end{figure}

As stated in (\ref{SOP}), SOP is another indicator for physical layer security system. In Fig.\ref{sop_beta} and Fig.~\ref{sop_beta_1}, we plot the SOP at different SNRs for $R_t$ = 2 bps/Hz and $\beta^2 \ne 1$ or $\beta^2 = 1$, respectively.
From Fig.~\ref{sop_beta} and ~\ref{sop_beta_1}, SOP of all TAS schemes decreases as the SNR increases for both cases of $\beta^2 \ne 1$ or $\beta^2 = 1$.
Owing to the same reason in previous paragraph, our proposed schemes almost have the same SOP as the conventional scheme for $\beta^2 = 1$, while they degrade a little in terms of SOP for $\beta \ne 1$.
The SOP of single antenna selection scheme is lower than that of two antenna selection scheme.

Observing Figs.~\ref{rate_beta}-\ref{sop_beta_1} in detail, we can discover that, for $\beta^2 \ne 1$, k-NN based TAS scheme achieves the best performance, while SVM and NB based schemes obtain relatively worse performance, in terms of both secrecy rate and SOP.
We analyze the reasons as follows. As stated previously, the KPI, secrecy rate, has a complicated coupling relationship with the channel gain vector, $\mathbf{h}$, which is difficult to be decoupled.
SVM is a deterministic classification scheme. When solving a complicated coupling relationship, SVM has a difficulty to extract feature vectors and construct precise hyper-planes, which result in some misclassification.
Similarly, for NB-based scheme, its high deviation makes the probabilistic estimation not accurate.
However, the k-NN based scheme relies on neighbor data sets and has low deviation with sufficient samples; therefore, it can construct better classification in this scenario.

\subsection{Computational Complexity}
In this subsection, we compare the computational complexity for our proposed ML-based schemes and conventional scheme. As stated in Section \ref{con sch}, $|\mathcal{L}|$ presents the number of selected antenna combinations, and $\left| \mathcal{L} \right| = \mathbb{C}_{N_\mathrm{S}}^{N_{\mathrm{T}}} = \frac{N_{\mathrm{S}}!}{N_{\mathrm{T}}! (N_{\mathrm{S}} - N_{\mathrm{T}})!} $. Let $N = N_\mathrm{S} + 1$. The selection complexity for SVM, NB, k-NN and conventional schemes are $\mathcal{O}(N^2)$, $\mathcal{O}(|\mathcal{L}|N+|\mathcal{L}|\log|\mathcal{L}|)$, $\mathcal{O}(N)$ and $\mathcal{O}(N+|\mathcal{L}|\log|\mathcal{L}|)$, respectively \cite{Joung2016Machine, He2018Transmit}. We can clearly see that the ML schemes is rather lower that of conventional schemes. It is because that the conventional TAS scheme requires to process the global search and comparison for each antenna combination. However, the complexity of ML-based schemes relies on the prediction complexity rather than the training complexity because the model training can be performed offline.

\subsection{Classification Performance}
Actually, the TAS is equivalent to a classification system with ML algorithm. In this subsection, we present the misclassification rate for single antenna selection by using the web representation \cite{Diri2008Visualization} for $\beta^2 = 1$ or $\beta^2 \ne 1$, $\mathrm{SNR}=15~\mathrm{dB}$. The value of each point in polygon denotes the misclassification rate of the corresponding channel index by $l \to \overline l $, where $l, \overline l \in \mathcal{L}$ and $l \ne \overline l $.
It can be seen that the misclassification rate for $\beta^2 \ne 1$ is higher than the rate for $\beta^2 = 1$. The critical reason is that, for $\beta^2 \ne 1$, the secrecy rate has a complicated couple in (\ref{R_s_case2}), which cause the ML algorithm cannot well decouple; as a result, classification performance is degraded.

\begin{figure}[ht]
    \centering
    \includegraphics[width=0.3\textwidth] {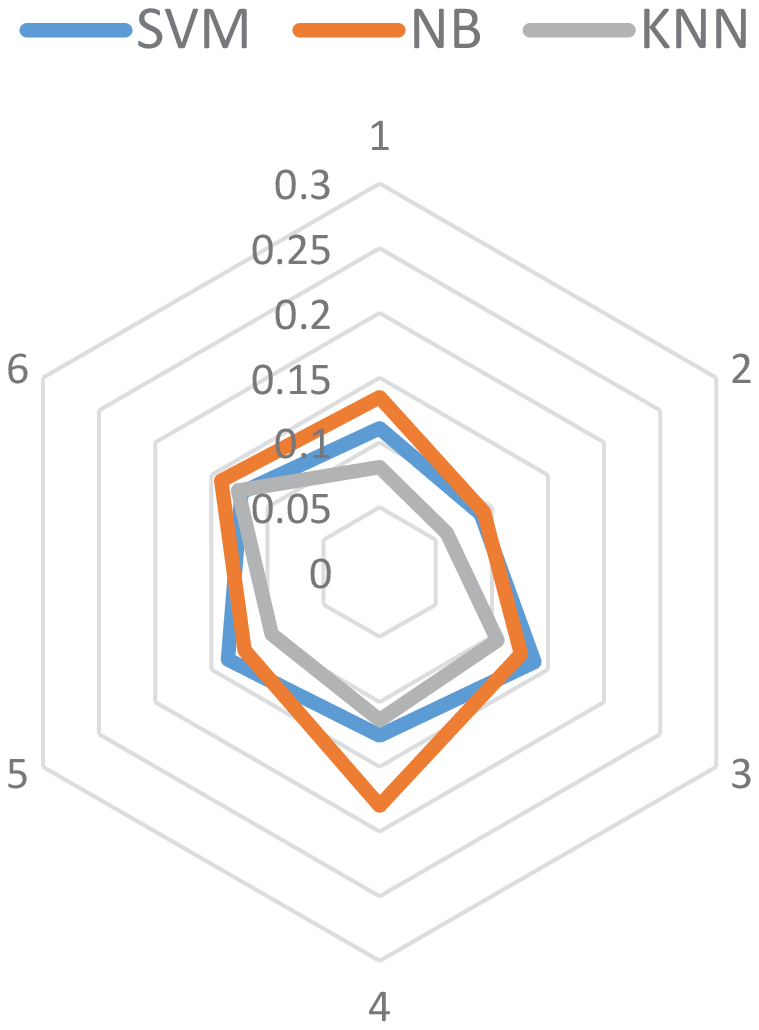}
    \caption{Misclassification rate for single antenna selection for $\beta^2 \ne 1$. }  \label{subfig:fig_nointer}
\end{figure}

\begin{figure}[ht]
    \centering
    \includegraphics[width=0.3\textwidth] {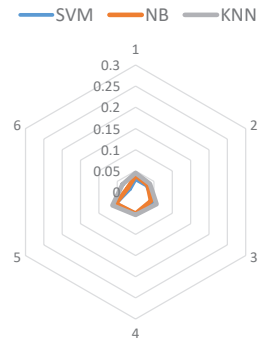}
    \caption{Misclassification rate for single antenna selection for $\beta^2 = 1$. }  \label{subfig:fig_snr}
\end{figure}

\section{conclusions}\label{conclusion}

In this paper, we applied ML algorithms for the antenna selection in untrusted relay networks to guarantee the performance as well as reducing the complexity. After pre-processing the training data sets, we constructed the SVM, NB and k-NN ML-based models to classify the CSI, which can be used to select the best antenna combination. The simulation results demonstrate that the ML-based TAS schemes can achieve almost the same performance as the conventional scheme. However, when amplification power constraint at the relay is considered, the complicated coupling relationship brings out some misclassification and thus degrades the achievable performance. For the case, we plan to introduce deep learning based scheme to decouple the relationship and improve the precision of TAS.

\bibliographystyle{IEEEtran}
\bibliography{ref}

\end{document}